\documentclass[aip,aapt,reprint,longbibliography]{revtex4-2}

\usepackage{natbib}
\usepackage{amssymb,amsmath}

\usepackage{epsfig}
\usepackage{bm}
\usepackage{color}
\usepackage{graphicx}
\usepackage{hyperref}

\usepackage{enumerate}

\usepackage{stmaryrd}

\newcommand\beq{\begin{equation}}
\newcommand\eeq{\end{equation}}
\newcommand\beqa{\begin{eqnarray}}
\newcommand\eeqa{\end{eqnarray}}
\newcommand{\nn}{\nonumber\\}
\def\bal#1\eal{\begin{align}#1\end{align}}

\newcommand{\dd}{\text{d}}
\newcommand{\ee}{\text{e}}

\newcommand{\ccdot}{\times}

\begin{document}

\title{The Newcomb--Benford law: Scale invariance and a simple Markov process based on it}

\author{Andrea Burgos}\email{anburgosr@alumnos.unex.es}

\author{Andr\'es Santos}\email{andres@unex.es}\affiliation{Departamento de F\'{\i}sica, Universidad de   Extremadura, 06006 Badajoz, Spain}

\date{\today}

\begin{abstract}
The Newcomb--Benford law, also known as the first-digit law, gives  the probability
distribution associated with the first digit of a dataset, so that, for example, the first significant digit has a
probability of $30.1$~\% of being $1$ and  $4.58$~\% of being $9$. This law can be extended to the
second and next significant digits.
This article presents an introduction to the discovery of the law, its derivation from the scale invariance property, as well as some applications and examples.
Additionally, a simple model of a Markov process inspired by  scale invariance  is proposed. Within this model, it is proved that the probability distribution  irreversibly converges to the Newcomb--Benford law, in analogy to the irreversible evolution toward equilibrium of physical systems in thermodynamics and statistical mechanics.

\end{abstract}

\maketitle

\section{Introduction}
\label{sec1}

In the late 19th century, an astronomer and mathematician visits his institution's library and consults a table of logarithms to perform certain astronomical calculations. As on previous occasions, he is struck by the fact that the first pages (those corresponding to numbers that start at $ 1$) are much more worn than the last ones (corresponding to numbers that start at $ 9$). Intrigued, this time he decides not to overlook the matter. He closes his eyes to concentrate, sketches a few calculations on a piece of paper, and finally smiles. He has found the answer and it turns out to be enormously simple and elegant.

A little over half a century later, a physicist and electrical engineer who is unaware of his predecessor's discovery, observes the same curious phenomenon on the pages of logarithm tables, and arrives at the same conclusion. Both have understood that, in a long list of records $\{r_n\}$ obtained from measurements or observations, the fraction $ p_d $ of records beginning with the significant digit $ d = 1,2, \ldots, 9 $ is not $ p_d = 1/9 $, as one might naively expect, but rather follows a logarithmic law. More specifically,
\beq
p_d=\log_{10}\left(1+\frac{1}{d}\right),\quad d=1,2,\ldots, 9.
\label{1}
\eeq
The numeric values of $ p_d $ are shown in the second column of Table \ref{table_LNB}. We see that the records that start with $ 1 $, $ 2 $, or $ 3 $ account for around $ 60 $~\% of the total, while the other six digits must settle for the remaining $ 40 $~\%.

\begin{table}
\caption{Probabilities for the first, second, third, and fourth significant digits.} \label{table_LNB}
\begin{ruledtabular}
\begin{tabular}{ c  c  c cc}
 Digit  & First&Second&Third & Fourth\\
 $d$&$p_d$&$p_d^{(2)}$&$p_d^{(3)}$&$p_d^{(4)}$\\
 \hline
 $ 0$& $\cdots$&$ 0.119\,68$&$ 0.101\,78$&$ 0.100\,18$\\
 $1$&$ 0.301\,03$&$ 0.113\,89$&$ 0.101\,38$&$ 0.100\,14$\\
 $2$&$ 0.176\,09$&$ 0.108\,82$&$ 0.100\,97$&$ 0.100\,10$\\
 $3$&$ 0.124\,94$&$ 0.104\,33$&$ 0.100\,57$&$ 0.100\,06$\\
 $4$&$ 0.096\,91$&$ 0.100\,31$&$ 0.100\,18$&$ 0.100\,02$\\
 $5$&$ 0.079\,18$&$ 0.096\,68$&$ 0.099\,79$&$ 0.099\,98$\\
 $6$&$ 0.066\,95$&$ 0.093\,37$&$ 0.099\,40$&$ 0.099\,94$\\
 $7$&$ 0.057\,99$&$ 0.090\,35$&$ 0.099\,02$&$ 0.099\,90$\\
 $8$&$ 0.051\,15$&$ 0.087\,57$&$ 0.098\,64$&$ 0.099\,86$\\
 $9$&$ 0.045\,76$&$ 0.084\,90$&$ 0.098\,27$&$ 0.099\,82$\\
\end{tabular}
\end{ruledtabular}
\end{table}

\begin{figure}
\includegraphics[width=5cm]{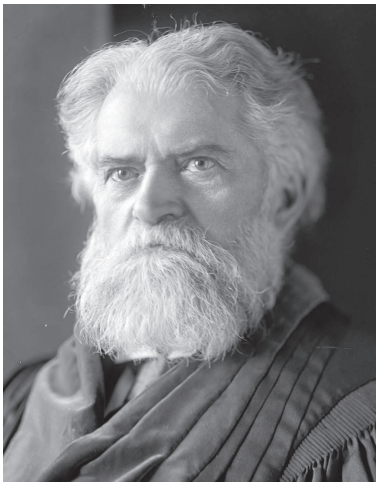}
\caption{\label{fig1} Simon Newcomb (1835--1909).}
\end{figure}
\begin{figure}
\includegraphics[width=5cm]{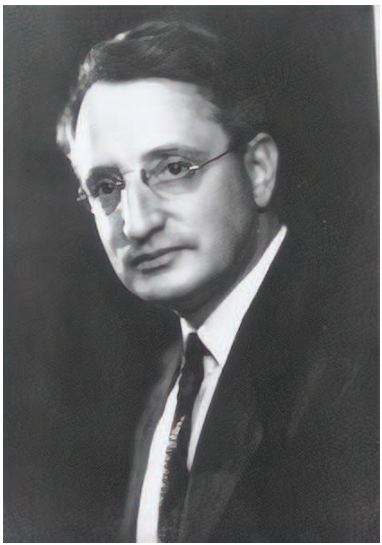}
\caption{\label{fig2} Frank Benford (1883--1948).}
\end{figure}

Our 19th century character is Simon Newcomb (Fig.\ \ref{fig1}) and he published his discovery in a modest two-page note.\cite{N81} The second character is Frank Benford (Fig.\ \ref{fig2}) and he wrote a $ 22 $-page article\cite{B38} in which, in addition to mathematically justifying Eq.\ \eqref{1}, he showed its validity in the analysis of more than $20\,000$ first digits taken from sources as varied as river areas, populations of American cities, physical constants, atomic and molecular weights, specific heats, numbers taken from newspapers or  Reader's Digest, postal addresses, \ldots, and the series $ n ^ {- 1} $, $ \sqrt{n} $, $ n^2 $, or $ n! $, among others, with $ n  \in \llbracket 1,100 \rrbracket $.

With such an overwhelming evidence, it is not surprising that Eq.\ \eqref{1} is usually known as \emph{Benford's law} (or first-digit law), even though it was found nearly sixty years earlier by Newcomb. This is  one more manifestation of the so-called Stigler's law, according to which no scientific discovery is named after the person who discovered it in the first place. In fact, as Stigler himself points out,\cite{S80} the law that bears his name was actually spelled out in a similar way twenty-three years earlier by the American sociologist Robert K. Merton. In order not to fall completely into Stigler's law, many authors refer to Eq.\ \eqref{1} as \emph{Newcomb--Benford's law} and that is the convention (by means of the acronym NBL) that we will follow in this article.

While the literature on the NBL in specialized journals is vast,\cite{BHR09} and several books on the topic are available,\cite{N12,BH15,*M15,*K15,*K19,*N20} the number of papers in general or science education journals is much scarcer.\cite{GF44,*FH45,*L86,*BK91,*P02,*TFGS07,*F09,*MA18,*L19,BMP93,H98}
In this paper, apart from revisiting the NBL and illustrating it with a few examples, we construct a Markov-chain model inspired by the invariance property of the NBL under the operation of a change of scale.

The remainder of the paper is organized as follows. The connection between the NBL (and some of its generalizations) with the property of scale invariance is formulated in Sec.\ \ref{sec2}. This is followed by a few examples in Sec.\ \ref{sec3}. The most original part is contained in Sec.\ \ref{sec4}, where our Markov-chain model is proposed and solved. Finally, the paper is closed in Sec.\ \ref{sec5} with some concluding remarks. For the interested reader, Appendices \ref{appA}--\ref{appC} contain the most technical and mathematical  parts of Secs. \ref{sec2} and \ref{sec4}.

\section{Origin of the law}
\label{sec2}
Often times, when one first speaks to a friend, relative, or even a colleague about the NBL, their first reaction is usually of skepticism. Why is the first digit not evenly distributed among the nine possible values? A simple argument shows that, if a robust distribution law exists, it cannot be the uniform distribution whatsoever.

\begin{figure}
\includegraphics[width=5cm]{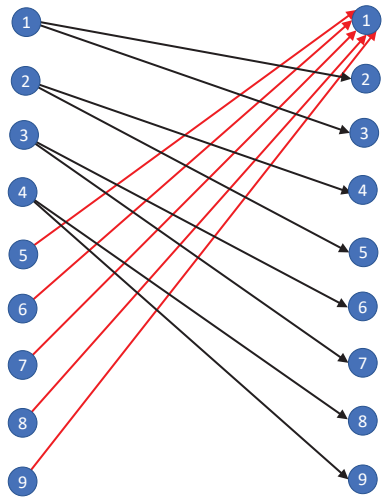}
\caption{\label{fig0} Diagram showing how the first digit changes when all the records of a dataset are doubled.}
\end{figure}

Imagine a long list of river lengths, mountain heights, and country surfaces, for example. It is possible that the lengths of the rivers are in km, the heights of the mountains in m, and the surfaces of countries in $ \text {km}^ 2 $, but they could also  be expressed in miles, feet, or acres, respectively. Will the distribution $ p_d $ depend on whether we use some units or others, or even if we mix them? It seems logical that the answer should be negative; that is,  the $ p_d $ distribution should be (statistically) independent of the units chosen; in other words,  it is expected that the $ p_d $ distribution is \emph{invariant under a change of scale}. The uniform distribution $ p_d = \frac{1}{9} $ obviously does not follow that property of invariance. Suppose we start from a dataset in which all the values of the first digit are equally represented. If we multiply all the records in the dataset by $ 2 $, we can see that those records that started before with $ 1 $, $2$, $3$, and $4$ then start with either $ 2 $ or $ 3 $, either $4$ or $5$, either $6$ or $7$, and either $8$ or $9$, respectively. On the other hand, all those that started with $ 5 $, $ 6 $, $ 7 $, $ 8 $, or $ 9 $ start now with $ 1$.  All those possibilities are schematically shown in Fig.\ \ref{fig0}.
Therefore, if $p_d = \frac{1}{9}$ initially, then $p_1 = \frac{5}{9}$ and $p_2 + p_3 = p_4 + p_5 = p_6 + p_7 = p_8 + p_9 =\frac{1}{9} $ after doubling all the records, thus destroying the initial uniformity.

Thus, the most identifying hallmark of the NBL is that it must be applied to records that have units or, as Newcomb himself writes,\cite{N81} ``As natural numbers occur in nature, they are to be considered as the ratios of quantities.''

While relatively formal mathematical proofs of the NBL can be found in the literature,\cite{P61,*F66,*R69,*R76,*D77,*N84,*H95b,*LSE00,*CLM19,*CM19,*V20,*BH20,H95a,BH11} in Appendix \ref{appA}, we present a sketch of a simpler, physicist-style  derivation of the law by imposing invariance under a change of scale.\cite{Weiss}

Equation \eqref{1} can be generalized beyond the first digit. The probability that the $m$ first digits of a record match the ordered string $(d_1, d_2, \ldots, d_m)$, where $d_1 \in \{1,2, \ldots, 9 \}$ and $d_i\in \{ 0,1,2, \ldots, 9 \}$ if $i\geq 2$, is given by (see Appendix \ref{appA})
\begin{equation}
\label{X2}
p_{d_1,d_2,\ldots, d_m}=\log_{10}\left[1+\left(\sum_{i=1}^m d_i\ccdot 10^{m-i}\right)^{-1}\right].
\end{equation}
As an example, the probability that the first three digits of a record form precisely the string $ (3,1,4) $ is $p_{3, 1, 4}= \log_{10} \left(1+ {1}/{314} \right)=0.001\,38$.

Once we have $ p_{d_1, d_2, \ldots, d_m} $, we can calculate the probability $ p_d ^{(m)} $ that the $ m $th digit is $ d $, regardless of the values of the preceding $ m-1 $ digits, just by summing for all possible values of those preceding $ m-1 $ digits,
\begin{equation}
\label{X7}
p_d^{(m)}=\sum_{d_1=1}^9\sum_{d_2=0}^9\cdots \sum_{d_{m-1}=0}^9 p_{d_1,d_2,\ldots, d_{m-1},d}.
\end{equation}
In Table \ref{table_LNB}, the law for the first digit, $ p_d $, is accompanied by the laws for the second, third, and fourth digits, obtained from Eqs.\ \eqref{X2} and \eqref{X7}. As the digit moves to the right, the probability distribution becomes less and less disparate.

In Fig.\ \ref{fig0} we saw that, when multiplying a dataset $ \{r_n \} $ by $ 2 $, part of the records that started with $ d = 1,2,3,4 $, specifically those that start with the first two digits $ (d, 0) $, $ (d, 1) $, $ (d, 2) $, $ (d, 3) $, or $ (d, 4) $, will start with $ 2d $, while the rest will start with $ 2d + 1$. Let us call $ \alpha_d $ the fraction of records that, initially starting with $ d = 1,2,3,4 $, start with $ 2d $ by doubling all the records. Thus,
\beq
\alpha_d=\frac{\sum_{d_2=0}^4 p_{d,d_2}}{p_d},\quad d=1,2,3,4.
\label{7}
\eeq
If the dataset fulfills the NBL, then one has
\begin{subequations}
\label{eq3.20-23}
\bal
\alpha_{1}=&\frac{\log_{10} \dfrac{3}{2}}{\log_{10} 2}\simeq 0.584\,96,
\quad \alpha_{2}=\frac{\log_{10}\dfrac{5}{4}}{\log_{10} \dfrac{3}{2}}\simeq 0.550\,34,
\label{eq3.20}
\\
\alpha_{3}=& \frac{\log_{10}\dfrac{7}{6}}{\log_{10}\dfrac{4}{3}}\simeq 0.535\,84,\quad
\alpha_{4}=\frac{\log_{10}\dfrac{9}{8}}{\log_{10}\dfrac{5}{4}}\simeq 0.527\,84. \label{eq3.23}
\eal
\end{subequations}
We will use these values later in Sec.\ \ref{sec4}.

\section{Applications and examples}
\label{sec3}

The applications and verifications of the NBL are numerous and cover topics as varied and prosaic as the study of the genome,\cite{HRJB02} atomic,\cite{P08} nuclear,\cite{BMP93,NR08,*NWR09} and particle\cite{SM09,*DD18} physics, astrophysics,\cite{MSPZ06,*SM10a,*AL14,*HF16,*SPP17,*JBR20} quantum correlations,\cite{CDSPSS16,*BMRBSS18,*BDSRSS18} toxic emissions,\cite{MH06} biophysics,\cite{SFSL20} medicine,\cite{CR16} dynamical systems,\cite{TBL00,*SCD01,*BBH04} distinction of chaos from noise,\cite{LFY15} statistical physics,\cite{SM10b} scientific citations,\cite{CC11,*AYS14} tax audits,\cite{NM09,N12} electoral\cite{CG07,*R14,*GA18} or scientific\cite{D07} frauds, gross domestic product,\cite{SAZ18,*AEKD19} stock market,\cite{H98,PTTV01} inflation data,\cite{MZDT19} climate change,\cite{LC19} world wide web,\cite{DMO06} internet traffic,\cite{AJ14} social networks,\cite{G15} textbook exercises,\cite{SID2015}  image processing,\cite{J01} religious activities,\cite{M12,*M14,*A14} dates of birth,\cite{AHI15} hydrology and geology,\cite{NM07,*STJ10,*GM12,*ACL17} fragmentation processes,\cite{K03,*BBCGIJMPRSST18} the first letters of words,\cite{YYKM18} or even COVID-19.\cite{LHJ20,*KY20} Other examples can be seen in the link of Ref.\ \onlinecite {testingBL}.
In this section, we will present some additional examples.

\begin{figure}
\includegraphics[width=8cm]{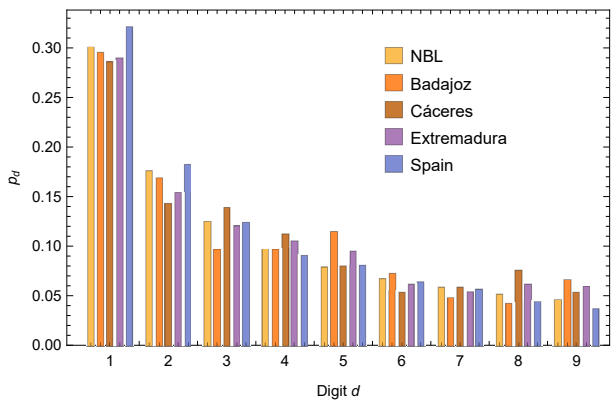}
\caption{\label{fig:muni} Histograms showing the distribution of the first digit for (from left to right at each digit $d$) the NBL and  the populations  of the municipalities of the provinces of Badajoz and C\'aceres, the region of Extremadura, and Spain.}
\end{figure}

Let us start with one of the situations that Benford himself studied in his classic paper:\cite{B38}  city populations. Using data from the Spanish National Institute of Statistics,\cite{INE} we have considered the 2019 population  of the $ 165 $ municipalities in the province of Badajoz (plus the total population of the province of Badajoz), of the $ 223 $ municipalities of the province of C\'aceres (plus the total population of the province of C\'aceres), and the total population of the $ 388 $ municipalities of the Spanish region of Extremadura,  which encompasses the provinces of Badajoz and C\'aceres, (plus the total populations of the provinces of Badajoz and C\'aceres). We have also considered the population (according to the $2016$ census) of the $ 8\,110 $ Spanish municipalities.
For all these lists of records, we have analyzed the frequency of those starting with $ d = 1,2, \ldots 9 $ and the results are compared in Fig.\ \ref{fig:muni}. There is a good general agreement between the population data and the NBL predictions. This is interesting, since it is not obvious that the distribution of the number of inhabitants of municipalities should be invariant under a change of scale.

\begin{figure}
\includegraphics[width=8cm]{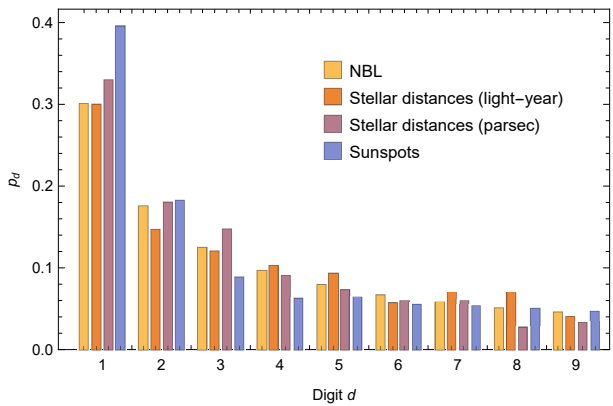}
\caption{\label{fig:astro} Histograms showing the distribution of the first digit for (from left to right at each digit $d$) the NBL, the distances to Earth in light-years and in parsecs from the brightest $ 300 $ stars, and the daily number of sunspots.}
\end{figure}

Let us now turn to two examples from astronomy. In the first one, we take the distance from Earth (in light-years and in parsecs) to the $ 300 $ brightest stars.\cite{Hipparcos}
In the second case, the data are the daily number of sunspots from $ 1818 $ to the present.\cite{sunspot} As seen in Fig.\ \ref{fig:astro}, the distances between our planet and the $300$ brightest stars agree reasonably well with the NBL, despite the fact that the list is not excessively long (only $ 300 $ data points) and that there are ``local'' deviations (for example, $ p_6 <p_7 $ in the two choices of units). This general agreement was to be expected, since the distribution of digits in distances (which are expressed in units) is a clear example of invariance under a change of scale. However, in the case of the daily number of sunspots, quantitative (although not qualitative) differences are observed with the NBL, especially in the $d = 1$, $ 3$, $ 4$, and $ 5$ cases. It should be noted that, although the series is rather long (more than $59\,000$ records, after excluding days without data or with zero spots), each record only has one, two, or three digits (the maximum number of sunspots was $ 528 $ and corresponded to August 26, 1870), thus producing a certain bias to records starting with $1$. Moreover, the daily number of sunspots cannot be expressed in units and therefore may not be scale-invariant.

\begin{figure}
\includegraphics[width=8cm]{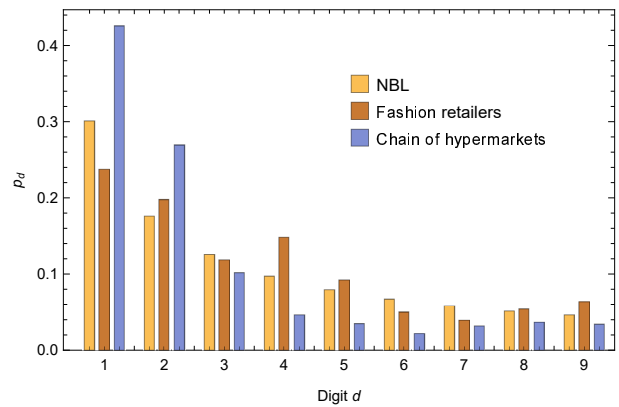}
\caption{\label{fig:prices} Histograms showing the distribution of the first digit for (from left to right at each digit $d$) the NBL and the prices of articles of a chain of fashion retailers and a chain of hypermarkets.}
\end{figure}

Finally, we have analyzed the prices of $1\,016$ items from a chain of fashion retailers \cite{Cortefiel} and of $1\,373$ products from a chain of hypermarkets \cite{Hipercor}. The results are shown in Fig.\ \ref{fig:prices}. In this case, the discrepancies with the NBL are more pronounced. Although the highest frequencies occur for $ d = 1 $ and $ d = 2 $, the observed values of $ p_d $ do not decrease monotonically with increasing $d$. In the case of the fashion retailers, we have $ p_4> p_3 $ and $ p_9> p_8> p_6> p_7 $; in the prices of the chain of hypermarkets, $ p_8> p_9> p_7> p_6 $. In principle, one might think that, since they can be expressed in euro, pound, dollar, peso, ruble, yen, \ldots, prices should satisfy the property of invariance under a change of scale inherent to the NBL. However, commercial and artificial pricing strategies must be superimposed on this invariance, which generates relevant deviations with respect to the NBL.

\section{A simple model of a Markov chain based on the scale invariance property of the Newcomb--Benford distribution}
\label{sec4}

\begin{figure}
\includegraphics[width=8cm]{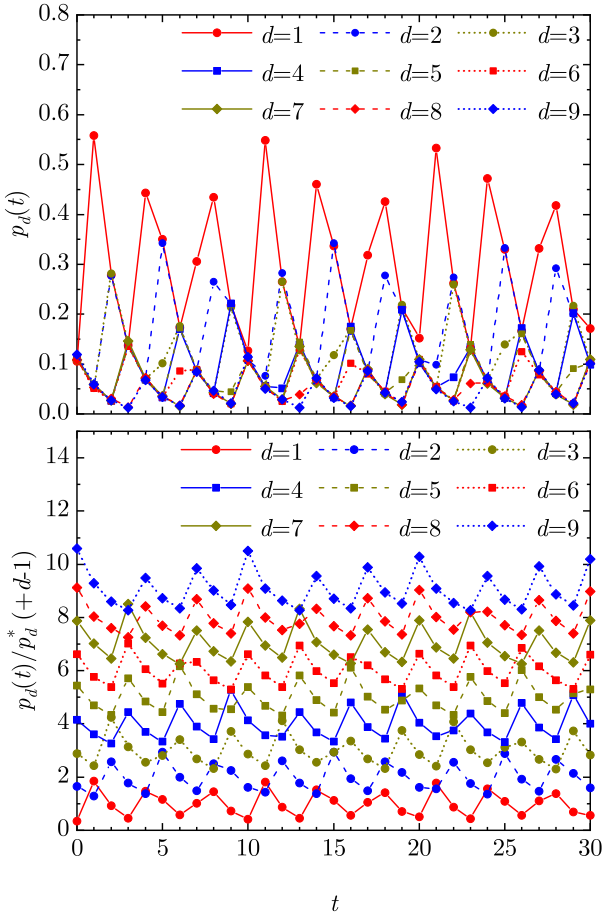}
\caption{\label{fig:true_uni} Evolution of the first-digit distribution $p_d(t)$ (top panel) and of the ratio $p_d(t)/p_d^*$ (bottom panel, where a vertical shift $d-1$ has been applied for better clarity),  $\{p_d^*\}$ being the NBL distribution, when starting from a dataset of $10^4$ random records uniformly distributed between $0$ and $1$ and doubling the records at each time step. Note the overlap of some of the points in the top panel.}
\end{figure}

\begin{figure}
\includegraphics[width=8cm]{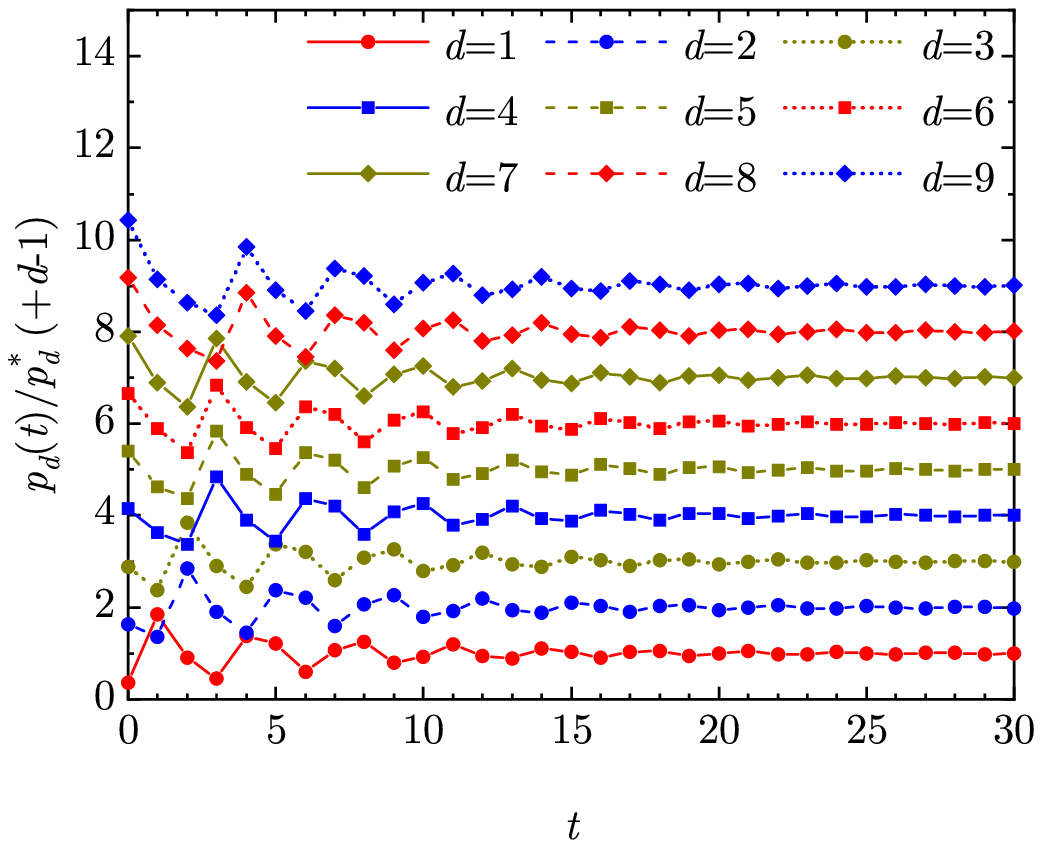}
\caption{\label{fig:uni} Evolution of the ratio $p_d(t)/p_d^*$ (where a vertical shift $d-1$ has been applied for better clarity), when starting from a uniform initial distribution $p_d(0)=\frac{1}{9}$, according to our Markov-chain model, Eq.\ \eqref{Markov}.}
\end{figure}

\begin{figure}
\includegraphics[width=8cm]{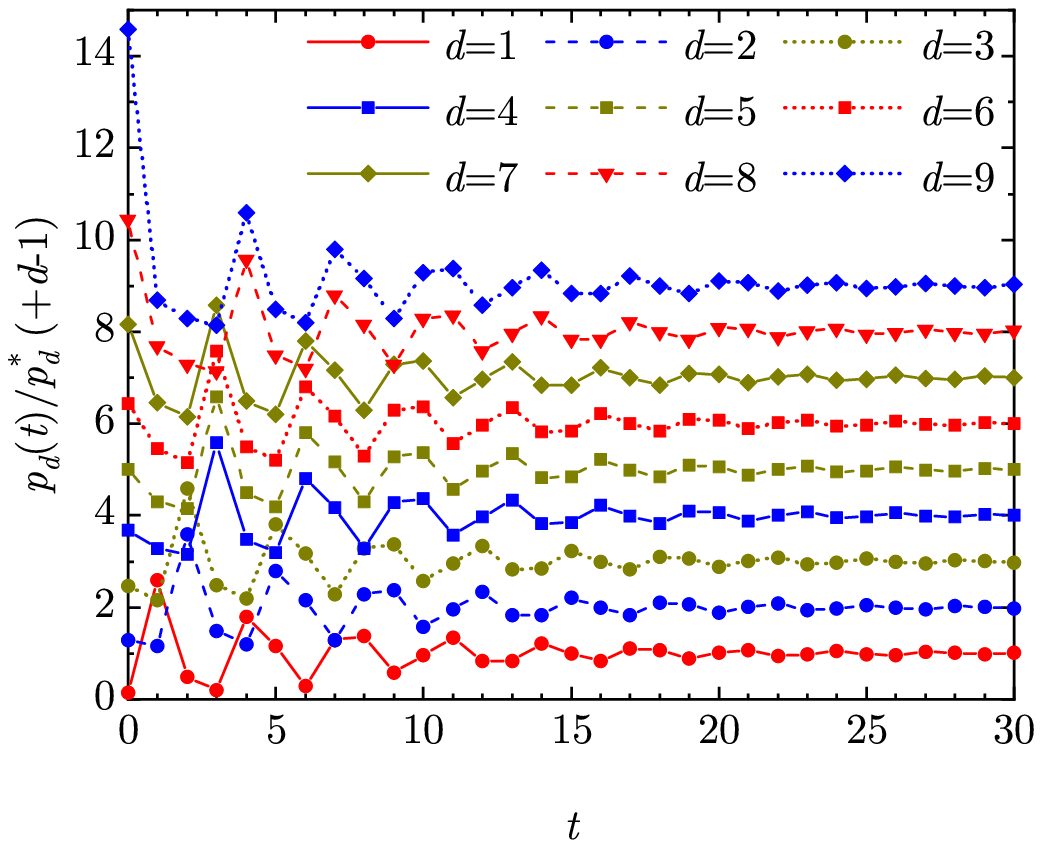}
\caption{\label{fig:inve} Evolution of  the ratio $p_d(t)/p_d^*$ (where a vertical shift $d-1$ has been applied for better clarity), when starting from an inverted initial distribution $p_d(0)=p_{10-d}^*$, according to our Markov-chain model, Eq.\ \eqref{Markov}.}
\end{figure}

\begin{figure}
\includegraphics[width=8cm]{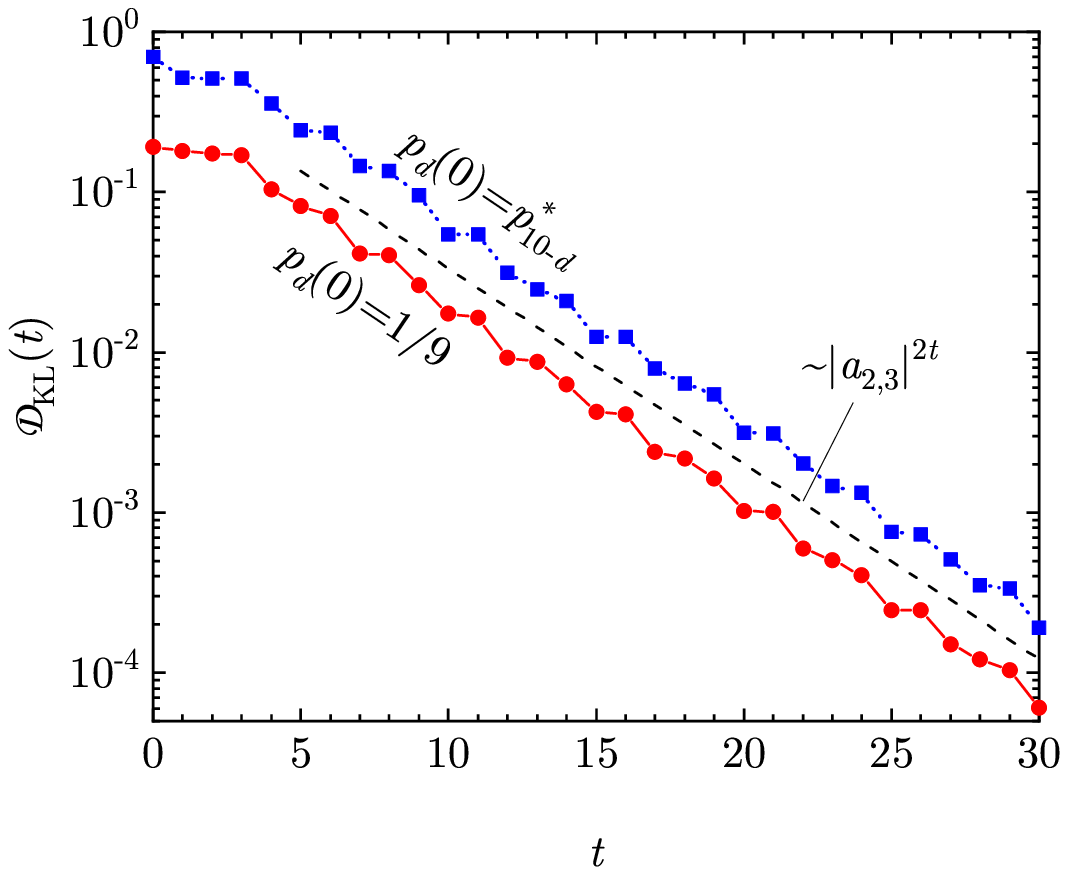}
\caption{\label{fig:KL} Evolution of the Kullback--Leibler divergence $\mathcal{D}_{\text{KL}} (t)$ (in logarithmic scale), starting from the uniform initial distribution $p_d(0)=\frac{1}{9}$ and from the inverted initial distribution $p_d(0)=p_{10-d}^*$, according to our Markov-chain model, Eq.\ \eqref{Markov}. The dashed line is proportional to $|a_{2,3}|^{2t}$ (see Appendix \ref{appC}).}
\end{figure}

As already said, NBL, Eq.\ \eqref{1}, is invariant under a change of scale; that is, if we start from a dataset $ \{r_n \} $ that fulfills the NBL and multiply all the records by a constant $ \lambda $, the resulting dataset $ \{\lambda r_n \} $ still fulfills the NBL.

It is tempting to conjecture that the NBL should be an \emph{attractor} of this process. This would mean that, if we started from an initial dataset $ \{r_n(0) \} $ that \emph{does not} fulfill the NBL and generated new sets $ \{r_n(t) \}=\{\lambda^t r_n(0) \} $ at times $t=1, 2,  \ldots$ by multiplying each successive  dataset by $\lambda $ (other than a fractional power of $10$), the first-digit distribution $\{p_d(t)\}$ of the generated sets would converge toward the NBL. However, this is not the case. As illustrated in Fig.\ \ref{fig:true_uni} for the case $\lambda=2$ and an initial dataset of $10^4$ random numbers, the time-dependent distribution $\{p_d(t)\}$ exhibits a quasiperiodic oscillatory pattern around the NBL distribution without any apparent amplitude attenuation. In fact, since $2^{10}=1\,024\simeq 10^3$,  the distribution nearly returns to the uniform initial one at times $t=10, 20, 30, \ldots$. This behavior is reminiscent of the Poincar\'e recurrence time in mechanical systems and the associated Zermelo paradox about irreversibility.\cite{S82}

In the transformation $\{r_n(t)\}\to \{r_n(t+1)=2r_n(t)\}$, the first-digit distribution changes, according to Fig.\ \ref{fig0}, as
\begin{subequations}
\label{3.1-3.8}
\begin{equation}
\label{3.1}
p_{1}(t+1)=p_{5}(t)+p_{6}(t)+p_{7}(t)+p_{8}(t)+p_{9}(t),
\end{equation}
\begin{equation}
\label{3.2}
p_{2}(t+1)=\alpha_{1} p_{1}(t),\quad
p_{3}(t+1)=(1-\alpha_{1}) p_{1} (t),
\end{equation}
\begin{equation}
\label{3.4}
p_{4}(t+1)= \alpha_{2} p_2(t),\quad
p_{5}(t+1)=(1-\alpha_{2}) p_{2}(t),
\end{equation}
\begin{equation}
\label{3.6}
p_{6}(t+1)=\alpha_{3} p_{3}(t),\quad
p_{7}(t+1)= (1-\alpha_{3}) p_{3}(t),
\end{equation}
\begin{equation}
\label{3.8}
p_{8}(t+1)= \alpha_{4} p_{4}(t), \quad
p_{9}(t+1)=(1-\alpha_{4})p_{4}(t),
\end{equation}
\end{subequations}
where the fractions $\alpha_d$ ($d=1,2,3,4$) are defined by Eq.\ \eqref{7}.
Note that, in general, the fractions $ \alpha_d $  depend on the distributions of the first digit, $p_{d}(t)$, and of the \emph{first two digits}, $p_{d,d_2}(t)$, of the dataset $ \{r_n (t) \} $ [see Eq.\ \eqref{7}]. As a consequence, (i) Eqs.\ \eqref{3.1-3.8} do not make a closed set and (ii) the parameters $ \alpha_d $ depend on $t$.

At this point, we construct a simplified dynamical model  in which the four  unknown and time-dependent parameters $ \alpha_d $ in Eqs.\ \eqref{3.1-3.8} are replaced by \emph{fixed} constants.
In matrix notation,
\begin{equation}
\label{Markov}
  \bm{p}(t+1)=\mathsf{W}\cdot \bm{p}(t),
\end{equation}
where ${\bm{p}}(t) = \left( p_{1}(t),p_{2}(t),\ldots,p_9(t)\right)^{\dagger} $ is a column vector ($\dagger$ denotes transposition) and
\begin{equation}
\label{W}
\mathsf{W}=
\begin{pmatrix}
0&0&0&0&1&1&1&1&1\\
\alpha_1&0&0&0&0&0&0&0&0\\
1-\alpha_1&0&0&0&0&0&0&0&0\\
0&\alpha_2&0&0&0&0&0&0&0&\\
0&1-\alpha_2&0&0&0&0&0&0&0&\\
0&0&\alpha_3&0&0&0&0&0&0&\\
0&0&1-\alpha_2&0&0&0&0&0&0&\\
0&0&0&\alpha_4&0&0&0&0&0&\\
0&0&0&1-\alpha_4&0&0&0&0&0&\\
\end{pmatrix}
\end{equation}
is a fixed square matrix.
Equation \eqref{Markov} fits  the canonical description of Markov chains,\cite{L84} where $\mathsf{W}$ is the so-called transition matrix, and $\{\alpha_d\}$ correspond to transition probabilities. In this way, we may forget about the meaning of $\{p_d(t)\}$ as the first-digit distribution of the dataset $\{r_n(t)\}$ and look at the  numerals $1,2,\ldots,9$ as labels of nine distinct internal states of a certain physical system which follow each other in the prescribed order sketched by Fig.\ \ref{fig0}.

Note that the matrix $\mathsf{W}$ is singular, that is, not invertible. This implies the \emph{irreversible} character of the transition $ \{p_d (t) \} \to \{p_d (t + 1) \} $. The stationary solution $\bm{p}^*$ to Eq.\ \eqref{Markov} satisfies the condition $\bm{p}^*=\mathsf{W}\cdot\bm{p}^*$.
Such a stationary solution will be an \emph{attractor} of our Markov process if $ \lim_ {t \to \infty} \bm{p} (t) = \bm{p}^{*} $  for \emph{any initial condition} $ \bm {p}(0) $. This property, along with the exact solution of Eq.\ \eqref{Markov},  is proved in Appendix \ref{appB}.
If  we choose for $\alpha_d$ the values given by Eqs.\ \eqref{eq3.20-23}, the stationary solution coincides exactly with the NBL, as could be expected. This will be the choice made in the rest of this section.

To illustrate the irreversible evolution of $\bm{p}(t) $ toward $\bm{p}^* $, we are going to consider two different initial conditions. First, we start from a uniform distribution, that is, $ p_d (0) = \frac{1}{9} $. The result is shown in Fig.\ \ref{fig:uni}, where we see that the evolution is oscillatory (see Appendix \ref{appB} for an explanation) and the oscillations are rapidly damped to attain the stationary solution. As a second example, we take an NBL inverted initial distribution, that is, $ p_d (0) = p_{10-d}^* $, so that, initially, state $ 9 $ is the most probable and  state $ 1 $ is the least probable. In this second example, as shown in Fig.\ \ref{fig:inve}, the initial oscillations are of greater amplitude but, as before, the stationary distribution is practically reached after a few iterations. Comparison between Figs.\ \ref{fig:true_uni} and \ref{fig:uni} shows that the main difference between our Markov model and the non-Markovian evolution of $\{p_d(t)\}$ in a real dataset is that the oscillation amplitudes are attenuated in the model and not in the original framework.

It seems convenient to characterize the evolution of the set of probabilities $ \{p_{d} (t) \}$ obeying the Markov process [Eq.\ \eqref{Markov}] toward the attractor $ \{p_ {d}^{*} \}$ by means of a single parameter that, in addition, evolves monotonically, thus representing the irreversibility of evolution.
It is expected that these properties are verified by the Kullback--Leibler divergence,\cite{KL51} which in our case is defined as
\begin{equation}
\label{3.15}
\mathcal{D}_{\text{KL}} (t) = \sum _{d=1} ^{9} p_{d} (t) \ln \frac{p_{d} (t)}{p_{d} ^{*}}.
\end{equation}
This quantity represents the opposite of the  Shannon entropy\cite{M99} of $\{p_d(t)\}$, except that it is measured  with respect to the stationary distribution $\{p_d^*\}$. In the context of our Markov-chain model, $\mathcal{D}_{\text{KL}}$ is related to information theory. On the other hand, the mathematical structure of $\mathcal{D}_{\text{KL}}$ can be extended to physical systems, thus  providing a statistical-mechanical basis to the thermodynamic entropy,\cite{M99,BN08} as exemplified below.

Figure \ref{fig:KL} shows the evolution of $ \mathcal {D}_{\text{KL}} (t) $ for the same initial conditions as in Figs.\ \ref{fig:uni} and \ref{fig:inve}. Both cases confirm the desired monotonic evolution of $\mathcal{D}_{\text{KL}} (t) $. Also, the asymptotic decay $\mathcal{D}_{\text{KL}} (t)\to 0 $ occurs essentially exponentially with a rate independent of the initial probability distribution. This is proved in Appendix \ref{appC}, as well as  the monotonicity property
\beq
\label{eq:H}
\mathcal{D}_{\text{KL}} (t+1)\leq\mathcal{D}_{\text{KL}} (t),
\eeq
with the equality not holding for two successive times unless $p_d(t)=p_d^*$, in which case $\mathcal{D}_{\text{KL}}=0$.

\begin{table*}
\caption{Analogy between a classical dilute gas (as a prototypical physical system) and a system described by the Markov chain, Eq.\ \eqref{Markov}. In the expression of the Maxwell--Boltzmann distribution $f_{\text{MB}}(\vec{v})$, $m $ is the mass of a molecule, $T$ is the gas temperature, and $k_B$ is the Boltzmann constant. In the Boltzmann equation, $J[\vec{v}|f(t),f(t)]$ is the collision operator.} \label{table:analogy}
\begin{ruledtabular}
\begin{tabular}{ l  c c}
 & Dilute gas&Markov chain\\
  \hline
Probability distribution&Velocity distribution function: $f(\vec{v},t)$&Probability of the internal state $d$: $p_d(t)$\\
  Normalization&$\displaystyle{\int \dd^3\vec{v}\,f(\vec{v},t)=1}$&$\displaystyle{\sum_{d=1}^9 p_d(t)=1}$\\
  Evolution equation&Boltzmann equation: $\displaystyle{\frac{\partial f(\vec{v},t)}{\partial t}=J[\vec{v}|f(t),f(t)]}$& $\displaystyle{\bm{p}(t+1)=\mathsf{W}\cdot\bm{p}(t)}$\\
    Equilibrium state&Maxwell--Boltzmann: $\displaystyle{f_{\text{MB}}(\vec{v})=\left(\frac{m}{2\pi k_BT}\right)^{3/2}\ee^{-mv^2/2k_BT}}$&NBL: $p_d^*=\log_{10}\left(1+d^{-1}\right)$\\
Entropy functional&$\displaystyle{S(t)=-\int \dd^3\vec{v}\,f(\vec{v},t)\ln\frac{f(\vec{v},t)}{f_{\text{MB}}(\vec{v})}}$&
$\displaystyle{S(t)=-\sum_{d=1}^9p_d(t)\ln\frac{p_d(t)}{p_d^*}}$\\
Irreversibility equation&$\displaystyle{\frac{\dd S(t)}{\dd t}\geq 0}$&$\displaystyle{S(t+1)-S(t)\geq 0}$\\
 \end{tabular}
\end{ruledtabular}
\end{table*}

Thus, we can say that the NBL in our Markov model plays a role analogous to the equilibrium state in thermodynamics and statistical mechanics. In this analogy, the ``entropy'' of the out-of-equilibrium system would be (except for a constant)  $ S = - \mathcal{D}_ {\text{KL}}  $, so that $ S $ increases irreversibly in the evolution toward equilibrium.

This analogy is put forward in Table \ref{table:analogy},  where we compare a system described by the Markov chain [Eq.\ \eqref{Markov}] to  a classical dilute gas as an example of a well-known physical system.
In both cases, a  statistical description is constructed by introducing the adequate  probability distribution: the {velocity} distribution function (gas) and the probability of the internal state $d$ (Markov chain); while $f(\vec{v},t)$ is continuous in both $\vec{v}$ and $t$,  $p_d(t)$ is discrete in $d$ and $t$. The  evolution equation for the probability distribution function is the Boltzmann equation (gas, under the molecular chaos ansatz\cite{CC70,*C88,*GS03,*K07,*B15}) and  Eq.\ \eqref{Markov} (Markov chain). The equilibrium state is characterized by  the Maxwell--Boltzmann distribution\cite{H65,*MB72,*LC07,*W13,*RMO17,*R19} $f_{\text{MB}}(\vec{v})$ (gas) and the NBL distribution $p_d^*$ (Markov chain).
In both classes of systems the nonequilibrium entropy functional $S(t)$ can be defined, within a constant, as the opposite  of the Kullback--Leibler divergence\cite{KL51,T20} of the equilibrium distribution from the time-dependent one. Boltzmann's celebrated $H$-theorem\cite{CC70,*C88,*GS03,*K07,*B15,P90,*B11,*R12} (gas) and the result presented  in Eq.\ \eqref{eq:H}  (Markov chain) show that the entropy $S(t)$ never decreases, irreversibly evolving toward its equilibrium value.

\section{Concluding remarks}
\label{sec5}

One of the main goals of this article was to show that, contrary to what might be initially thought, the first significant digit $d$ of a dataset extracted from measurements or observations of the real world is not evenly distributed among the nine possible values, but typically the frequency is higher for $ d = 1 $ and decreases as $ d $ increases. The NBL, Eq.\ \eqref{1}, gives a mathematical expression to this empirical fact, although it does not always need to be rigorously verified due to statistical fluctuations (as happens with populations in Fig.\ \ref{fig:muni} and with distances in Fig.\ \ref{fig:astro}), limitations of the sample (as in the sunspot case in Fig.\ \ref{fig:astro}), or an artificial bias (as in the case of prices of articles in Fig.\ \ref{fig:prices}). It is to be expected that, except for unavoidable statistical fluctuations, the law is fulfilled in  datasets in which quantities are measured in units, so that the distribution of the first digit is independent of the units chosen due to its invariance under a change of scale. More generally, the NBL is satisfied when the mantissa of the logarithms of the considered data is uniformly distributed. That makes lists not directly related to physical quantities, such as Fibonacci numbers or powers of $ 2$,  also satisfy the NBL.

Gaining inspiration from the scale invariance property of the NBL, we have constructed a Markov-chain model for a nine-state system that evolves in time according to the scheme depicted in Fig.\ \ref{fig0}. The initial-value model can be exactly solved, the solution showing an irreversible relaxation toward the NBL probability distribution. Moreover, we have proved that the associated (relative)  entropy monotonically increases, in analogy with the second law of thermodynamics in physical systems.

Until the $1970$s  (which is when scientific pocket calculators began to be used), physicists used tables of logarithms (or their application in slide rules) for small everyday scientific calculations.
Those calculations are nowadays performed on pocket calculators, cellular phones, or personal computers with a wide variety of existing mathematical programs. Since the data that are
manipulated in physics are extracted from ``real'' situations, such as experiments, models, physical constants, \ldots, we can conclude, as a tribute to Newcomb and Benford and their logarithmic tables, that the keyboard button  $ 1 $ will be the one that presents the greatest wear and tear, while that of $ 9 $ will be the least used.

\begin{acknowledgments}

A.S. acknowledges financial support from Grant No.\ FIS2016-76359-P/AEI/10.13039/501100011033 and the Junta de Extremadura (Spain) through Grant No.\ GR18079, both partially financed by Fondo Europeo de Desarrollo Regional funds.
\end{acknowledgments}

\appendix
\section{Derivation of the NBL and some generalizations}
\label{appA}

Consider a long list of records $\{r_n\}$ that, without loss of generality for the matter at hand, we will assume positive. Each record can be written in the form $r_n = x_n \ccdot 10^{k_n}$, where $ k_n $ is an integer and $x_n \in [1,10) $ is the \emph{significand}. Obviously, it is the distribution of the significand that is relevant for the NBL. The significand $ x_n $ is directly related to the \emph{mantissa} $\mu_n$ of the decimal logarithm of $r_n$, that is, $\log_ {10} r_n = k_n + \mu_n$, where $\mu_n = \log_{10} x_n \in [0,1)$.

Let $P_x (x) \dd x$ be the probability that the significand lies between $x$ and $x + \dd x$, so that the normalization condition is $\int_1^{10} \dd x \, P_x (x) = 1$. The probability that the first significant digit of any record  is $d$ is then given by the integral
\beq
\label{3}
p_d=\int_d^{d+1}\dd x\, P_x(x).
\eeq
More generally, if we consider an ordered string $(d_1, d_2, \ldots, d_m)$ made up of the first $ m $ digits, where $ d_1 \in \{1,2, \ldots, 9 \} $ and $ d_i \in \{ 0,1,2, \ldots, 9 \} $ if $ i\geq 2$, it is obvious that the records whose $ m $ first digits match the string $ (d_1, d_2, \ldots, d_m) $ will be those whose significand $ x $ is greater than or equal to $ d_1 + d_2 \ccdot 10^{- 1} + \cdots + d_m \ccdot 10^{-(m-1)} $ and less than $ d_1 + d_2 \ccdot 10^{- 1} + \cdots + (d_m + 1) \ccdot 10^{- (m-1) } $. Consequently,
\begin{equation}
\label{X2b}
p_{d_1,d_2,\ldots, d_m}=\int_{\sum_{i=1}^m d_i\times 10^{-(i-1)}}^{10^{-(m-1)}+\sum_{i=1}^m d_i\times 10^{-(i-1)}}\dd x\, P_x(x).
\end{equation}

If the distribution $P_x(x)$ is actually invariant under a change of scale, that means that $P_x (\lambda x) = f(\lambda) P_x (x)$ with arbitrary $\lambda$. Taking into account the normalization condition in the form $\int_\lambda^{10 \lambda} \dd (\lambda x) \, P_x (\lambda x) = 1$, it follows that necessarily $f(\lambda ) = \lambda^{- 1}$, that is, $P_x (\lambda x) = \lambda ^ {-1} P_x (x)$. Differentiating both sides of the equation with respect to $\lambda $ and then taking $\lambda = 1$, we easily obtain $xP_x'(x) = - P_x (x)$, which, according to Euler's theorem, implies that $P_x (x)$ is a homogeneous function of degree $-1$, that is, $P_x (x) \propto x^{- 1}$. The constant of proportionality is obtained from the normalization condition, finally yielding
\beq
P_x(x)=\frac{x^{-1}}{\ln 10},\quad 0\leq x<10.
\label{4}
\eeq
This is the \emph{unique} distribution of significands that is invariant under a change of scale. From Eq.\ \eqref {4}, and applying Eqs.\ \eqref {3} and \eqref{X2b}, it is straightforward to obtain  NBL, Eq.\ \eqref{1}, and its generalization, Eq.\ \eqref{X2}.
As a consistency test of Eq.\ \eqref{X2}, it is easy to check that
\bal
\label{X3}
p_{d_1,d_2,\ldots, d_{m-1}}=&\sum_{d_m=0}^9 p_{d_1,d_2,\ldots, d_m}\nn
=&\log_{10}\left[1+\left(\sum_{i=1}^{m-1} d_i\ccdot 10^{m-1-i}\right)^{-1}\right].
\eal

It is interesting to note that the inverse law for the significand, Eq.\ \eqref{4}, implies a uniform law for the mantissa (and vice versa). Let $P_\mu(\mu) \dd\mu $ be the probability that the mantissa lies between $\mu$ and $\mu + \dd \mu$. Since $ P_\mu (\mu) \dd \mu = P_x (x) \dd x $ and $ \dd \mu = (x^{-1} / \ln 10) \dd x $, Eq.\ \eqref {4} gives us $P_\mu (\mu) = 1$. In Newcomb's words,\cite {N81} ``The law of probability of the occurrence of numbers is such that all mantiss{\ae} of their logarithms are equally probable.'' An immediate consequence is that, if $\mu $ is a random variable uniformly distributed between $ 0 $ and $ 1 $, then the random variable $ x =10 ^\mu $ fulfills the NBL. This is in fact an easy way to generate a list of random records matching the NBL.

There are deterministic series that also satisfy the NBL. Suppose the series $ \{r_n = a \alpha^n + b \beta^n, n = 1,2, \ldots \} $, where $a$, $b$, $\alpha$, and $\beta$ are real numbers with $ a > 0 $, $ \alpha >  \beta \geq 0 $, and $ \log_{10} \alpha  = \text{irrational}$.\cite{BH11} In that case, $\lim_{n \to \infty} \log_{10} r_n = n \log_{10}  \alpha  + \log_{10}  a  $ has a uniformly distributed mantissa, so $ \{r_n\} $ satisfies the NBL. That includes, for example, the series $ \{2^n \} $, $ \{3^n \}$, and $ \{F_n \} $, where $F_n = \frac{1}{\sqrt{5 }} \left[\varphi^n- (-\varphi^{- 1})^n \right] $ are the Fibonacci numbers,  $ \varphi \equiv \frac{1}{2} \left (1+ \sqrt{5} \right) $ being the golden ratio. Similarly, the series $ \{n!\} $ also verifies the law.\cite{BH11}

Another important property is that if a list $ \{r_n \} $ fulfills the NBL, so does the list $ \{r_n^a \} $, $a$ being a real number. Indeed, if $\log_ {10} r_n = k_n + \mu_n $, the mantissa $ \mu_n $ being uniformly distributed, then the mantissa of $ \log_{10} r_n^a = a (k_n + \mu_n) $ is also evenly distributed. This is directly related to the fact that the NBL is not only invariant under a change of scale but also under \emph{base change},\cite{H95a} as would be expected, given the artificial character of the decimal base. To see it, let us assume a base $ b $ and build the list $ \{r_n^a \} $, with $ a = \log_{10} b $, from a list $ \{r_n \} $ that fulfills the NBL. In that case, $ r_n^a = y_n \ccdot b^{k_n} $, where $ y_n = x_n^a \in [1, b) $ is the significand of $r_n^a $ in the base $ b $. The probability distribution $ P_y(y) $ is related to the distribution $ P_x(x) $ through the equation $P_y (y) \dd y = P_x (x) \dd x $, so that Eq.\ \eqref {4} leads to
\beq
P_y(y)=\frac{y^{-1}}{\ln b},\quad 0\leq y<b.
\label{5}
\eeq
Therefore,  NBL  in an arbitrary base $ b $ takes the form
\beq
p_d=\log_{b}\left(1+\frac{1}{d}\right),\quad d=1,2,\ldots, b-1.
\label{6}
\eeq

\section{Exact solution of the Markov-chain model}
\label{appB}
Note first that Eqs.\ \eqref{3.1-3.8} verify the normalization condition $ \sum_ {d = 1}^{9} p_{d} (t + 1) = \sum_ {d = 1}^{9} p_{d} (t) = 1 $. Therefore, only eight of the probabilities $ \{p_d, d = 1,2, \ldots, 9 \} $ are linearly independent, so we can eliminate one of them. If we choose $ p_ {9} = 1- \sum_ {d = 1}^{8} p_{d} $, Eq.\ \eqref{3.1} gives us
$p_{1}(t+1)=1- p_{1}(t)-p_{2}(t)-p_{3}(t)-p_{4}(t)$. Although it is not strictly necessary, it is mathematically convenient to split the column vector $\bm{p}(t)$ into the subsets ${\bm{p}}_{\text{I}}(t) \equiv \left( p_{1}(t),p_{2}(t),p_{3}(t),p_{4}(t)\right)^{\dagger} $,
 ${\bm{p}}_{\text{II}}(t) \equiv \left( p_{5}(t),p_{6}(t),p_{7}(t),p_{8}(t)\right)^{\dagger} $, and $p_9(t)$. As a consequence, the matrix equation \eqref{Markov} can be decomposed into
\begin{equation}
\label{3.11}
{\bm{p}}_{\text{I}}(t+1)= \bm{q}+ \mathsf{A} \cdot {\bm{p}}_{\text{I}}(t),\quad
{\bm{p}}_{\text{II}}(t+1)= \mathsf{B} \cdot {\bm{p}}_{\text{I}}(t),
\end{equation}
where  $\bm{q}=(1,0,0,0)^{\dagger}$ and
 \begin{subequations}
\begin{equation}
\label{3.12}
\mathsf{A}=
\begin{pmatrix}
-1 &-1&-1&-1\\
\alpha_{1}&0&0&0\\
1-\alpha_{1} &0&0&0\\
0 &\alpha_2&0&0
\end{pmatrix},
\end{equation}
\begin{equation}
\label{3.12B}
\mathsf{B}=
\begin{pmatrix}
0 &1-\alpha_2&0&0\\
0&0&\alpha_{3}&0\\
0&0&1-\alpha_{3} &0\\
0 &0&0&\alpha_4
\end{pmatrix}.
\end{equation}
\end{subequations}
In this way, one deals with two independent $4\times 4$ matrices instead of the $9\times 9$ transition matrix introduced in Eq.\ \eqref{W}.
Moreover, only the equation for the projected vector $\bm{p}_{\text{I}}$ in Eq.\ \eqref{3.11} needs to be solved. Once the solution is obtained, the solution for the complementary projected vector $\bm{p}_{\text{II}}$ is straightforward.

By recursion, it is easy to check that  the solution to the initial-value problem posed by Eq.\ \eqref{3.11} is
\begin{subequations}
\label{3.14B}
\begin{eqnarray}
\label{3.14}
{\bm{p}_{\text{I}}}(t)&=& \sum _{n=0} ^{t-1} \mathsf{A}^{n} \cdot \bm{q}+ \mathsf{A}^{t} \cdot {\bm{p}_{\text{I}}}(0)\nonumber \\
 &=& (\mathsf{I}-\mathsf{A}^t)\cdot \bm{p}_{\text{I}}^{*}+ \mathsf{A}^{t} \cdot {\bm{p}_{\text{I}}}(0),
\end{eqnarray}
\beq
\label{3.14bis}
\bm{p}_{\text{II}}(t)=\mathsf{B}\cdot(\mathsf{I}-\mathsf{A}^{t-1})\cdot \bm{p}_{\text{I}}^{*}+ \mathsf{B}\cdot\mathsf{A}^{t-1} \cdot {\bm{p}_{\text{I}}}(0),
\eeq
\end{subequations}
where $\mathsf{I}$ is the identity matrix and
\begin{subequations}
\label{3.13B}
\beq
\label{3.13}
\bm{p}_{\text{I}}^{*}=\left(\mathsf{I}-\mathsf{A} \right)^{-1} \cdot \bm{q}
=\frac{1}{3+\alpha_1\alpha_2}
\begin{pmatrix}
  1\\
  \alpha_1\\
  1-\alpha_1\\
  \alpha_1\alpha_2
\end{pmatrix},
\eeq
\beq
\bm{p}_{\text{II}}^{*}=\mathsf{B} \cdot \bm{p}_{\text{I}}^{*}
=\frac{1}{3+\alpha_1\alpha_2}
\begin{pmatrix}
  \alpha_1(1-\alpha_2)\\
  (1-\alpha_1)\alpha_3\\
  (1-\alpha_1)(1-\alpha_3)\\
  \alpha_1\alpha_2\alpha_4
\end{pmatrix},
\eeq
\end{subequations}
is the \emph{unique} stationary solution. {}From the normalization condition, one has $p_9^*=\alpha_1\alpha_2(1-\alpha_4)/(3+\alpha_1\alpha_2)$.
Note that, as seen from Eqs.\ \eqref{3.14B}, the initial values $\bm{p}_{\text{II}} (0) $ do not influence the evolution of either $\bm{p}_{\text{I}} (t) $ or $\bm{p}_{\text{II}} (t) $.

The stationary solution will be an \emph{attractor} if $ \lim_ {t \to \infty} \bm{p}_{\text{I}} (t) = \bm{p}_{\text{I}}^{*} $ and $ \lim_ {t \to \infty} \bm{p}_{\text{II}} (t) = \bm{p}_{\text{II}}^{*} $ for any initial condition $ \bm {p}_{\text{I}} (0) $. According to Eqs.\ \eqref{3.14B}, this implies $ \lim_ {t \to \infty} \mathsf{A}^{t} = 0 $.

To check the above condition, let us obtain the four eigenvalues $ \{a_i, i = 0,1,2,3 \} $ of $ \mathsf {A} $. It is easy to see that the characteristic equation is $ a (\alpha_1 \alpha_2 + a + a ^ 2 + a ^ 3) = 0 $. Therefore, the eigenvalues are $ a_0 = 0 $ and
\begin{subequations}
\label{17ab}
\beq
a_1=-\frac{1}{3}\left(1+\frac{2}{\beta}-\beta\right),
\eeq
\beq
a_{2,3}=-\frac{1}{3}\left(1-\frac{1\pm\imath\sqrt{3}}{\beta}+\frac{1\mp\imath\sqrt{3}}{2}\beta\right),
\eeq
\end{subequations}
where $\imath$ is the imaginary unit and
\beq
\beta\equiv \left[\frac{3}{2}\left(\frac{7}{3}-9\alpha_1\alpha_2+\sqrt{9-42\alpha_1\alpha_2+81\alpha_1^2\alpha_2^2}\right)\right]^{1/3}.
\eeq
Consequently,
\beq
\mathsf{A}^t=\mathsf{U}\cdot \mathsf{D}^t\cdot\mathsf{U}^{-1},\quad t=1,2,\ldots,
\eeq
where
\beq
\mathsf{U}=\begin{pmatrix}
0&\frac{a_1^2}{\alpha_1\alpha_2}&\frac{a_2^2}{\alpha_1\alpha_2}&\frac{a_3^2}{\alpha_1\alpha_2}\\
0&\frac{a_1}{\alpha_2}&\frac{a_2}{\alpha_2}&\frac{a_3}{\alpha_2}\\
-1&\frac{(1-\alpha_1)a_1}{\alpha_1\alpha_2}&\frac{(1-\alpha_1)a_2}{\alpha_1\alpha_2}&\frac{(1-\alpha_1)a_3}{\alpha_1\alpha_2}\\
1&1&1&1
\end{pmatrix},
\eeq
\beq
\mathsf{D}^t=
\begin{pmatrix}
0&0&0&0\\
0&a_1^t&0&0\\
0&0&a_2^t&0\\
0&0&0&a_3^t
\end{pmatrix}
,\quad t=1,2,\ldots.
\eeq
From Eqs.\ \eqref{17ab} it can be verified that $ | a_1 | <| a_ {2,3} | <1 $ for $ 0 <\alpha_1 \alpha_2 <1 $, so that $ \lim_{t \to \infty} a_i^t = 0$ for $i=1,2,3$. Therefore, $ \lim_{t \to \infty} \mathsf{D}^t = 0 $ and hence $ \lim_{t \to \infty} \mathsf{A}^t = 0 $. This proves the  character of the stationary distribution $\bm {p}^*$ as an attractor of the dynamics.
Moreover, since both the real eigenvalue ($ a_1 $) and the real part of the complex eigenvalues ($ a_{2,3} $) are negative, the evolution of each $p_d(t)$ to $p_d^*$ is \emph{oscillatory}, as can be observed in Figs.\ \ref{fig:uni} and \ref{fig:inve}.
Note that, in the analysis above, the eigenvalues $0$ (double), $1-\alpha_3$, and $\alpha_4$ of the matrix $\mathsf{B}$ are not needed.

If we choose $ \alpha_d = \frac{1}{2} $, then Eqs.\ \eqref{3.13B} provide the stationary solution $p_1^*=\frac{4}{13}\simeq 0.308$, $p_2=p_3=\frac{2}{13}\simeq 0.154$, $p_4=p_5=p_6=p_7=\frac{1}{13}\simeq 0.077$, $p_8=p_9=\frac{1}{26}\simeq 0.038$. These values are not too different from those of the true NBL, as can be seen from Table \ref{table_LNB}. On the other hand, the most natural choice is provided by  Eqs.\ \eqref{eq3.20-23}, in which case the stationary solution coincides exactly with the NBL.

\section{Properties of the Kullback--Leibler divergence in the Markov model}
\label{appC}
The Kullback--Leibler divergence is defined by Eq.\ \eqref{3.15}. In order to analyze its asymptotic  decay, let us consider times that are long enough so that the deviations $ \delta p_d (t) \equiv p_d (t) -p_d^* $ can be considered small. In this regime, we can expand Eq.\ \eqref{3.15} in a power series and retain the dominant terms. The result is
\begin{equation}
\label{3.15B}
\mathcal{D}_{\text{KL}} (t) \approx \frac{1}{2}\sum _{d=1} ^{9} \frac{\left[\delta p_{d} (t)\right]^2}{p_{d} ^{*}}.
\end{equation}
On the other hand, for times long enough, $ |a_1 |^t \ll |a_{2,3} |^t $ (note that $ | a_1 | = 0.4261 $ and $ | a_{2,3} | = 0.8692 $), so that, according to Eqs.\ \eqref{3.14B}, $ \delta p_d (t) \sim | a_{2,3} |^t $. Thus,
\beq
\mathcal{D}_{\text{KL}} (t) \sim | a_{2,3} |^ {2t} = 10^{2t \log_{10} | a_{ 2,3} |} .
\eeq
This asymptotic behavior is represented in Fig.\ \ref{fig:KL}.

Finally, let us prove Eq.\ \eqref{eq:H}. According to Eq.\ \eqref{3.15},
\bal
\label{C3}
\mathcal{D}_{\text{KL}} (t+1)=&p_1(t+1)\ln\frac{p_1(t+1)}{p_1^*}\nn
&+\sum_{d=1}^4 \left[p_{2d}(t+1)\ln\frac{p_{2d}(t+1)}{p_{2d}^*}\right.\nn
&\left.+p_{2d+1}(t+1)\ln\frac{p_{2d+1}(t+1)}{p_{2d+1}^*}\right].
\eal
Evolution equations \eqref{3.1-3.8} and the stationarity condition $\bm{p}^*=\mathsf{W}\cdot\bm{p}^*$ can be rewritten as
\begin{subequations}
\label{rewr}
\beq
p_1(t+1)=\sum_{d=5}^9p_d(t),\quad p_1^*=\sum_{d=5}^9p_d^*,
\eeq
\beq
\begin{Bmatrix}
p_{2d}(t+1)\\
p_{2d+1}(t+1)
\end{Bmatrix}
=\begin{Bmatrix}
\alpha_d\\
1-\alpha_d
\end{Bmatrix}
p_d(t),\quad d=1,2,3,4,
\eeq
\beq
\begin{Bmatrix}
p_{2d}^*\\
p_{2d+1}^*
\end{Bmatrix}
=\begin{Bmatrix}
\alpha_d\\
1-\alpha_d
\end{Bmatrix}
p_d^*,\quad d=1,2,3,4.
\eeq
\end{subequations}
Inserting Eqs.\ \eqref{rewr} into Eq.\ \eqref{C3} one gets
\beq
\label{3.16}
\mathcal{D}_{\text{KL}} (t+1) =\sum_{d=5}^9 p_d(t)\ln\frac{\sum_{d'=5}^9 p_{d'}(t)}{\sum_{d'=5}^9 p_{d'}^*}
+\sum_{d=1}^4p_d(t)\ln\frac{p_d(t)}{p_{d}^*}.
\eeq
Therefore,
\bal
\Delta \mathcal{D}_{\text{KL}}(t+1)\equiv &
\mathcal{D}_{\text{KL}} (t+1)-\mathcal{D}_{\text{KL}} (t)\nn=&
-\sum_{d=5}^9 p_d(t)\ln\frac{p_d(t)\sum_{d'=5}^9 p_{d'}^*}{p_d^*\sum_{d'=5}^9 p_{d'}(t)}.
\eal
Given the stationary values $\{p_d^*,d=5,\ldots,9\}$, the difference $\Delta \mathcal{D}_{\text{KL}}(t+1)$ is a function of the $5$ probabilities $\{p_d(t),d=5,\ldots,9\}$. To find the maximum value of $\Delta \mathcal{D}_{\text{KL}}(t+1)$, we take the derivative
\beq
\frac{\partial \Delta \mathcal{D}_{\text{KL}}(t+1)}{\partial p_d(t)}=-\ln\frac{p_d(t)\sum_{d'=5}^9 p_{d'}^*}{p_d^*\sum_{d'=5}^9 p_{d'}(t)}.
\eeq
The unique solution to the extremum conditions ${\partial \Delta \mathcal{D}_{\text{KL}}(t+1)}/{\partial p_d(t)}=0$ is $p_d(t)=\gamma p_d^*$ ($d=5,\ldots,9$), where $0<\gamma<1/\sum_{d=5}^9 p_d^*$ is arbitrary. In such a case, $\Delta \mathcal{D}_{\text{KL}}(t+1)=0$. To see that this is actually a maximum value, suppose, for instance, that $p_d(t)=0$ except if $d=d_0$ (with $d_0=5,\ldots, 9$). In that case, it is easy to find $\Delta \mathcal{D}_{\text{KL}}(t+1)=p_{d_0}(t)\ln\left(p_{d_0}^*/\sum_{d=5}^9 p_{d}^*\right)<0$.
We then conclude that $\Delta \mathcal{D}_{\text{KL}}(t+1)\leq 0$, i.e., we obtain Eq.\ \eqref{eq:H},
the equality holding only if $p_d(t)=\gamma p_d^*$ ($d=5,\ldots,9$). Note that, even though $\Delta \mathcal{D}_{\text{KL}}(t+1)= 0$ if  $p_d(t)/p_d^*=\text{constant}$ for $d=5,\ldots,9$, this proportionality property is broken down at $t+1$ unless $\gamma=1$. As a result, $\Delta \mathcal{D}_{\text{KL}}(t+2)< 0$, even though $\Delta \mathcal{D}_{\text{KL}}(t+1)= 0$, except in the stationary solution.

%

\end{document}